\documentclass{sig-alternate-2013}
\usepackage{amsmath}
\usepackage{times}
\usepackage{helvet}
\usepackage{courier}
\usepackage{graphicx}
\usepackage{color}

\permission{Copyright is held by the International World Wide Web Conference Committee (IW3C2). IW3C2 reserves the right to provide a hyperlink to the author's site if the Material is used in electronic media.}
\conferenceinfo{WWW 2015 Companion,}{May 18--22, 2015, Florence, Italy.} 
\copyrightetc{ACM \the\acmcopyr}
\crdata{978-1-4503-3473-0/15/05. \\
http://dx.doi.org/10.1145/2740908.2743058}

\begin{document}

\title{Deep Feelings: A Massive Cross-Lingual Study on the Relation between Emotions and Virality}

\numberofauthors{2}
\author{
\alignauthor
Marco Guerini\\
       \affaddr{Trento Rise}\\
       \affaddr{Via Sommarive 18}\\
       \affaddr{Trento, Italy}\\
       \email{marco.guerini@trentorise.eu}
\alignauthor
Jacopo Staiano\\
       \affaddr{LIP6, UPMC - Sorbonne Universit\'{e}s}\\
       \affaddr{and FBK MobS Lab}\\
       \affaddr{4 place Jussieu, Paris, France}\\
       \email{jacopo.staiano@lip6.fr}
}

\maketitle

\newcommand{\jacopo}[1]{\textcolor{red}{#1}}

\begin{abstract}
This article provides a comprehensive investigation on the relations between virality of news articles and the emotions they are found to evoke. Virality, in our view, is a phenomenon with many facets, i.e. under this generic term several different effects of persuasive communication are comprised. 
By exploiting a high-coverage and bilingual corpus of documents containing metrics of their spread on social networks as well as a massive affective annotation provided by readers, we present a thorough analysis of the interplay between evoked emotions and viral facets. 

We highlight and discuss our findings in light of a cross-lingual approach: while we discover differences in evoked emotions and corresponding viral effects, we 
provide preliminary evidence of a generalized explanatory model rooted in the deep structure of emotions: the Valence-Arousal-Dominance (VAD) circumplex. We find that viral facets appear to be consistently affected by particular VAD configurations, and these configurations indicate a clear connection with distinct phenomena underlying persuasive communication.
\end{abstract}




\section{Introduction}

The mass-adoption of social networking sites and the wide integration of sharing widgets on popular websites have paved the way for quantitative research efforts tackling the 
relations between content and virality. Very recently, a different kind of widgets is working its way through the online space: designed to take the 
\emph{affective} pulse of the websites visitors, such small interfaces allow people to explicitly tag their emotional states as they browse the web.

Although the adoption of the latter is not yet prominent, we found two very popular news outlets, one in English and one in Italian, embedding similar interfaces for affective 
feedback in each article page. We thus became interested in investigating the relation of this newly available \emph{affective} data with virality of news articles in a 
cross-lingual experimental setting. The two online websites which made this research possible are: 
\begin{enumerate}
 \item Rappler (\texttt{rappler.com}), an English-written ``social news'' portal, 
which makes extensive use of its distinctive \emph{Mood Meter} feature;
 \item the online version of the Italian newspaper Corriere della Sera (\texttt{corriere.it}), one of the most 
popular daily newspapers in Italy.
\end{enumerate}


In our previous work~\cite{staiano2014depeche}, we have shown the impressive value of such reader-provided affective data: we used Rappler data to automatically build an emotion lexicon and a 
system for automatic affective analysis of texts. 
In the evaluation results we outperformed existing systems even using a na\"{\i}ve approach, thanks to the quality and coverage of the human annotation of evoked emotions. In this work, we turn to analyze such data \emph{along} with indices of 
virality in order to derive insights on the relations between the emotions evoked by textual content and its diffusion and engagement.

Early works on emotions and virality have already been published~\cite{virality}, paving the way for this novel line of research. Still, a few limitations of~\cite{virality}, which this paper attempts to overcome, can be identified in (i), the use of a small sample of articles ($\sim$7,000) with only a subset manually annotated by three annotators ($\sim$2,500 documents); (ii), the authors only consider one virality index; and (iii), they only hint to a possible explanatory role of deep constituents of emotions. 

In this paper we leverage a large corpus of news articles (ten times bigger than \cite{virality}) with a massive crowd-sourced voluntary annotation of the emotion each article evokes in readers (more than 1.5 millions votes), so to:
\begin{itemize}
  \item compare several viral phenomena at once and understand if they are consistently affected by emotions and if they are affected in the same way;
 \item compare emotion and virality in a cross-lingual experimental setting;
 \item investigate the effect of deep constituents of emotions in viral phenomena.
\end{itemize}

Our findings show that, while the relations between emotions and virality seem to vary across cultures, their deeper constituents 
(i.e. the Valence, Arousal, and Dominance components, as defined in the VAD circumplex model of affect we adopt) show consistency among all indices of virality we account for, providing a generalized model for their interplay. More specifically, viral indices are coherently affected by particular VAD configurations, and these configurations point to a clear connection with general phenomena underlying different viral facets. 
These results are relevant not only for social science researchers interested in understanding the factors behind virality phenomena, but also for marketing and industry people as they can be very valuable in contexts such as content marketing and native advertising~\cite{hackley2014advertising}.

This paper is structured as follows: the following section provides the reader with a brief review of recent research efforts on virality, social media and emotion studies; 
in Section~\ref{sec:DS} we describe the data collection procedure followed; Sections~\ref{sec:emo} and~\ref{sec:vad} report our analyses and findings, while in
Section~\ref{sec:r_square} we provide a comparison with~\cite{virality}. Finally, we take stock of our work in Section~\ref{sec:concl}.

\section{Related Work}
\label{sec:relwork}
In this section we provide a short review of research efforts focused on (i) understanding content virality on Social Media and (ii) the role of emotions in Social Media.

\textbf{Content Virality.} Several researchers have studied information flow, community building and similar processes using Social Networking sites 
as a reference~\cite{credibility,digging,caseStudyOnComments,contagionDigg}. However, the great 
majority focused on network-related features without taking into account the actual content spreading within the network~\cite{voting}.
A hybrid approach focusing on both product characteristics and network related features is presented in~\cite{aral2011creating}: 
the authors study the effect of passive-broadcast and active-personalized notifications embedded in an application 
aimed at fostering word of mouth.

Recently, the relation between content characteristics and virality has begun to be investigated, especially 
with regard to textual content. In~\cite{opinion}, for instance,
features derived from sentiment analysis of comments are used to predict the popularity of stories. 
The relevant work in~\cite{danescu2012you} measures a different form of content 
spreading, by analyzing which are the features of a movie quote that make it ``memorable" online. 
Another approach to content virality, somehow complementary to the previous one, is presented in~\cite{simmons2011memes}, 
where the authors investigate which modification dynamics make a meme spread from one person to another 
(as compared with movie quotes which spread remaining exactly the same).
Louis and Nenkova~\cite{louis2013makes} focused on influential scientific articles in newspapers, considering characteristics such as readability, description vividness, use of unusual words 
and affective content, comparing high quality articles (NYT articles appearing in ``The Best American Science Writing" anthology) against typical NYT articles.  

Moreover, the work presented in~\cite{borghol2012untold} investigates how differences in textual description affect the spread of content-controlled videos.
In~\cite{lakkaraju2013s}, the authors focus on the act of resubmissions (i.e., content that is submitted multiple times with multiple titles to multiple different online communities) to understand the extent to which each factor influences the success of a content. In~\cite{tan2014effect} it is investigated how content spreads in an online community by pinpointing the effect of wording in terms of content informativeness, generality and affect.
Finally, Althoff et al.~\cite{althoff2014ask} developed a model that can predict the success of requests for a free pizza gifted from the Reddit community, 
using high level textual features such as politeness, reciprocity, narrative and gratitude.

More recently, some works have tried to investigate how different textual contents give rise to different reactions 
in the audience: the work presented in~\cite{marco:carlo:gozde:ICWSM-11} correlates several viral 
phenomena with the wording of a post, while in~\cite{guerini2012linguistic} it is shown that specific content features variations (like the readability level of an abstract) differentiate among virality level of downloads, 
bookmarking, and citations. Similarly, Shuai et al.~\cite{shuai2012scientific} studied scientific articles in terms of downloads, Twitter mentions, and early citations in the scholarly records, trying to understand how these virality indices correlate among them.
Finally, also in the realm of visual content it has been shown that different image characteristics can give rise to different viral phenomena~\cite{guerini2013exploring}. Following this line of research, we study the effects of emotions considering several audience reactions to find out whether there are peculiar emotional characteristics of a news article that give rise to different viral reactions. 

\textbf{Emotions in Social Media}. Kramer et al.~\cite{kramer2014experimental} have shown, via a massive experiment on Facebook, that emotional states can be transferred to others via emotional contagion, leading people to experience the same emotions without their awareness. The experiment included reducing the amount of emotional content in the News Feed of a user: when positive expressions were reduced, people produced fewer positive posts and more negative posts; when negative expressions were reduced, the opposite pattern occurred. While in~\cite{kramer2014experimental} the authors manipulated News Feed content, in our study we simply analyze existing and publicly available content voluntarly annotated by readers. 

In~\cite{hochreiter2014role}, the authors focused on the role of emotions in the context of word-of-mouth marketing, using a dataset of Google+ posts: their analyses show consistency with~\cite{virality}, with increase in \textsc{anger} linked to higher likelihood of reshares, while the opposite trend holds for \textsc{sadness}. Fan et al.~\cite{fan2013anger} looked at diffusion patterns on the very popular chinese platform Weibo, which shares many features with Twitter, and again found a similar result: angry posts appear to spread at a significantly faster rate, while sad posts do so at a significant lower rate.

Furthermore, the work presented in~\cite{hansen2011good} comes closer to the focus of the present paper by hypothesizing that negative news content is more likely to be retweeted, while for non-news tweets positive sentiments support virality. To test this hypothesis the authors analyze three corpora and give evidence that negative sentiment enhances virality in the news segment, but not in the non-news segment. Their conclusion is that
the relation between affect and virality is more complex than expected based on the findings of~\cite{virality}.

In~\cite{querciamood}, popular and influential users on twitter are linguistically analized according to their tweets. The initial hypothesis is that a Twitter account cannot be simply traced back to the graph properties of the network within which it is embedded, but also depends on the personality and emotions of the human being behind it. The reported findings suggest that popular users tend to use positive emotions while influential users lean towards negative ones. 

Finally, the work presented in~\cite{virality} uses \textit{New York Times} articles to examine the relationship between 
emotions evoked by the content and virality, using semi-automated sentiment analysis to quantify the affectivity 
and emotionality of each article. Results suggest a strong relationship between affect and virality; still, the 
virality metric considered is interesting but very limited: it only consists of how many people emailed the article.
We consider this work as a starting point for our research and for comparison of results: we will discuss it throughout the paper. 

\textbf{Human, crowd-sourced, voluntary, large scale annotations.} These distinguishing characteristics of the affective data employed in our analyses mark the significant difference between this article and previous works.
While previously mentioned research efforts have resorted to computational linguistics classifiers to automatically annotate textual content, the data we crawled (from publicly available websites) allows us to leverage affective annotations voluntarily provided by the news article readers.

\section{Dataset Collection}
\label{sec:DS}

\begin{figure}[ht!]
\centering
\includegraphics[width=50mm]{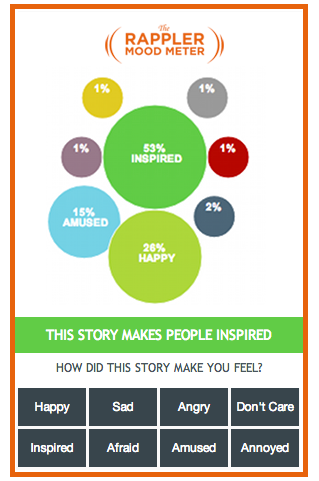}
\caption{Rappler's Mood Meter.}
\label{fig:moodmeter}
\end{figure}

\begin{figure}[ht!]
\centering
\includegraphics[width=\columnwidth]{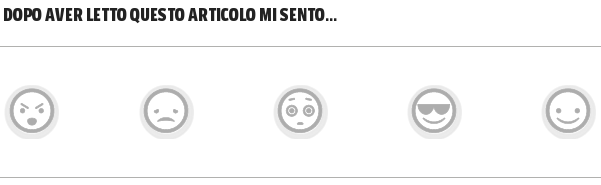}
\caption{The emoticon-based interface on \texttt{corriere.it} articles. The sentence translates to ``After reading this article, I feel..''}.
\label{fig:corriere}
\end{figure}

The ``social-news'' website \texttt{rappler.com} embeds a small interface, called \emph{Mood Meter}, in every article it publishes. Such interface allows the readers to express 
with a simple click their emotional reaction to the story they are reading. The percentages of votes obtained by each emotion are also visualized by the interface, which is depicted 
in Figure~\ref{fig:moodmeter}.
Similarly, the online version of a very popular Italian newspaper (\emph{Corriere della Sera}) has recently adopted a similar approach, based on 
\emph{emoticons}, to sense the emotional states of its readers, as shown in Figure~\ref{fig:corriere}\footnote{As we are forced to adopt a crawling approach, we cannot have any control on the layout of the widgets shown above. Thus, we cannot rule out the possibility of biases arising from the (fixed) order used to present readers with affective labels and emoticons. Still, the results reported in~\cite{staiano2014depeche} -- obtained thanks to data crawled with the same strategy, indicate that such bias, if present, is negligible.}.

Following previous work presented in~\cite{staiano2014depeche},
we harvested a total of 53,226 news articles from \texttt{rappler.com}, and 12,437 articles from \texttt{corriere.it}.
Our crawler was programmed to retrieve 
articles at least a month old, in order to allow the indicators to settle and thus not to penalize the most recent articles. The articles span over roughly one year for Rappler, and nine months for Corriere della Sera.
For the scope of this paper, the main difference between the two mechanisms lies in the fact that, while it is possible to extract absolute votes from \texttt{corriere.it}, along 
with percentage values, 
\texttt{rappler.com} only exposes the latter. In particular, for the bulk of articles crawled from \texttt{corriere.it}, our data includes a total of 320,697 votes 
for the five emotional dimensions available. 

 Although we have no means to verify the actual absolute number of votes collected by the Mood Meter, we can provide a very conservative estimate for it: by computing the lower common denominator over the percentages of affective votes obtained by a Rappler article, we can derive the minimum number of votes needed to obtain them, compounding to a total of 1,145,543 votes over the entire Rappler dataset. For comparison, consider that the same conservative estimate for the \texttt{corriere.it} data amounts to 210,113 -- less than two thirds of the actual value depicted above.

Thus, the datasets used in this work comprise more than 65,000 news articles and more than 1.5 million annotations.

Since the sets of emotions accounted for in the two websites differ both in size (\texttt{rappler.com} allows to tag eight affective dimensions, whereas five are available on \texttt{corriere.it}) and, 
although slightly, in semantics (being in two different languages), we need to proceed to map the subset available from the latter to the former, as reported in 
Table~\ref{tab:corriere_mapping}. 
 
\begin{table} [!htb] 	
	\begin{center} 	 	
		{
			\begin{tabular}{ll} 		
				\hline 
 \texttt{corriere.it} label & maps to\\
 \hline 
\textsc{Triste} & \textsc{Sad} \\
\textsc{Divertito} & \textsc{Amused} \\
\textsc{Soddisfatto} & \textsc{Happy*} \\
\textsc{Preoccupato} & \textsc{Afraid*} \\
\textsc{Indignato} & \textsc{Annoyed*} \\
\hline 
			\end{tabular} 		
		} 		 	
	\end{center}	 	
	\caption{Mapping of original emotion labels from \texttt{corriere.it} to those present in \texttt{rappler.com}. * denotes appropriate mapping, altough the English label is not the primary translation.} 	
	\label{tab:corriere_mapping} 
\end{table}

In Table~\ref{tab:percentage-votes} we report the mean percentage of votes for each emotional dimension on the two corpora: it can be seen that \textsc{happiness} has the highest percentage of votes by a large margin in comparison to the other dimensions in \texttt{rappler.com}, while it is very closely followed by \textsc{indignato/annoyed} in the \texttt{corriere.it} data.

\begin{table} [!htb] 	
	\begin{center} 	 	
		{ 		
			\begin{tabular}{lrr} 		
				\hline 
 & Rappler$_{\mu}$ & Corriere$_{\mu}$\\
 \hline 
\textsc{Afraid} & .05 & .09 \\
\textsc{Dont\_Care} & .05 & - \\
\textsc{Amused} & .11 & .11  \\
\textsc{Happy} & .31 & .29  \\
\textsc{Angry} & .11 & - \\
\textsc{Inspired} & .11 & - \\
\textsc{Annoyed} & .06 & .25 \\
\textsc{Sad} & .12 & .10 \\
\hline 
			\end{tabular} 		
		} 		 	
	\end{center}	 	
	\caption{Mean vote percentages obtained by emotions in Rappler and Corriere.} 	
	\label{tab:percentage-votes} 
\end{table}

\begin{table*}[!hbt]
\centering
	{\scriptsize
	\begin{minipage}[b]{.30\textwidth}
	\begin{tabular}{rrrr}
	\multicolumn{4}{c}{Rappler Comments} \\
  	\hline
 	& Estimate & Std. Err& Sigf. \\ 
  	\hline
 	 ANNOYED & .61 & .03  & *** \\ 
  	 ANGRY & .54 & .02 & *** \\ 
  	 INSPIRED & .23 & .02  & *** \\ 
  	 DONT\_CARE & .18 & .04  & *** \\ 
 	 AMUSED & .13 & .02  & *** \\ 
 	 HAPPY & .10 & .02  & *** \\ 
 	 SAD & .07 & .02  & *** \\ 
  	 AFRAID & -.03 & .03  & $\dagger$ \\ 
  	 \hline
 	\end{tabular}
		\end{minipage} 
		\begin{minipage}[b]{.30\textwidth}
	\begin{tabular}{rrrr}
	\multicolumn{4}{c}{Rappler Tweets} \\
  	\hline
	 & Estimate & Std. Err& Sigf. \\ 
  	\hline
  	INSPIRED & .52 & .02 & *** \\ 
	ANGRY & .32 & .02 & *** \\ 
	ANNOYED & .32 & .03 & *** \\ 
	HAPPY & .31 & .02 & *** \\
	DONT\_CARE & .30 & .04 & *** \\ 
	SAD & .24 & .02 & *** \\ 
	AMUSED & .21 & .02 & *** \\  	
	AFRAID & .19 & .03 & *** \\ 
   	\hline
	\end{tabular}
		\end{minipage} 	
		\begin{minipage}[b]{.30\textwidth}
	\centering
	\begin{tabular}{rrrr}
	\multicolumn{4}{c}{Rappler G$+$} \\
  	\hline
 	& Estimate & Std. Err& Sigf. \\ 
  	\hline
	INSPIRED & .55 & .02 & *** \\ 
  	ANGRY & .25 & .02 & *** \\ 
   	HAPPY & .24 & .02 & *** \\ 
	AMUSED & .21 & .02 & *** \\ 
  	ANNOYED & .20 & .03 & *** \\ 
  	SAD & .16 & .02 & *** \\ 
  	AFRAID & .14 & .03 & *** \\ 
	DONT\_CARE & .13 & .04 & *** \\ 
   	\hline
	\end{tabular}
	\end{minipage} 
	
	\caption{Emotions impact on \texttt{rappler.com} articles viral indices. ***: $p<$.001; **: $p<$.01; *: $p<$.05; $\dagger$: not significant -- this legend applies to all tables in this paper.}
	\label{tab:rappler_res}
	}

\end{table*}


\begin{table*}[!hbt]
\centering
	{\scriptsize
	\begin{minipage}[b]{.30\textwidth}
\begin{tabular}{rrrr}
\multicolumn{4}{c}{Corriere Comments} \\
  \hline
 & Estimate & Std. Err& Sigf. \\ 
  \hline
  ANNOYED & 1.11 & .04 & *** \\ 
  HAPPY & .40 & .04 & *** \\ 
  AFRAID & .19 & .06 & ** \\ 
  AMUSED & .13 & .06 & * \\ 
  SAD & .12 & .05  & * \\   
   \hline
\end{tabular}
		\end{minipage} 	
		\begin{minipage}[b]{.30\textwidth}
\begin{tabular}{rrrrr}
\multicolumn{4}{c}{Corriere Tweets} \\
  \hline
 & Estimate & Std. Err& Sigf. \\ 
  \hline
  ANNOYED & .33 & .03 & *** \\ 
  SAD & .20 & .05 & *** \\ 
  HAPPY & .14 & .03 & *** \\ 
  AFRAID & .06 & .05 & $\dagger$ \\ 
  AMUSED & .02 & .05 & $\dagger$ \\ 
   \hline
\end{tabular}
		\end{minipage} 	
		\begin{minipage}[b]{.30\textwidth}
\begin{tabular}{rrrr}
\multicolumn{4}{c}{Corriere G$+$} \\
  \hline
 & Estimate & Std. Err & Sigf. \\ 
  \hline
  SAD & .57 & .05 & *** \\ 
  ANNOYED & .39 & .03 & *** \\ 
  AMUSED & .34 & .05 & *** \\ 
  AFRAID & .33 & .05 & *** \\ 
  HAPPY & .27 & .03 & *** \\ 
   \hline
\end{tabular}
		\end{minipage} 	
		}
		\caption{Emotions impact on \texttt{corriere.it} articles viral indices.}
	\label{tab:corriere_res}
\end{table*}

The statistics on \texttt{rappler.com} emotional data confirm the trend noted in~\cite{staiano2014depeche} about \textsc{happiness} predominance, for which several explanations may be hypothesized: from cultural characteristics, 
to a bias in the dataset itself -- as it might contain mainly `positive' news, through psychological phenomena leading people to express more positive moods on social 
networks~\cite{de2012not,querciamood,vittengl1998time}. 

Studies on other English datasets, e.g. on LiveJournal posts~\cite{strapparava2008learning}, have in the past 
noted predominance of the happy mood.
Conversely, no previous studies have dealt at this scale with Italian language, and it will be worth investigating in future works what factors may influence the trend shown for 
emotional dimensions in the \texttt{corriere.it} data.

Turning to the statistics of the various viral indices available for the two datasets, we provide a summary in Table~\ref{tab:viral_counts}.

\begin{table} [!htb] 	
	\begin{center} 	 	
		{	
			\begin{tabular}{lrr} 		
				\hline 
 				Viral Index & Rappler$_{\mu}$ & Corriere$_{\mu}$\\
 				\hline 		
				\textsc{Comments} & 4.11 & 83.22\\
				\textsc{Threads} & 2.81 &	- \\
				\textsc{G$+$ shares} & .91 & 3.79 \\
				\textsc{Twitter shares} & 32.33 & 40.65 \\
				\textsc{Facebook shares} & - & 502.93 \\				
				\hline			
			\end{tabular} 	
		} 		 	
	\end{center}	 	
	\caption{Mean figures for virality indices.}
	\label{tab:viral_counts} 
\end{table}  

From now on, we will only consider the intersection of the two sets of indices.

\subsection{Narrow- and Broad- casting}
In the literature, \emph{broadcasting} refers to the act of communicating or transmitting a content to numerous recipients simultaneously, over a communication network; on the contrary, \emph{narrowcasting} has traditionally been understood as the dissemination of information to a narrow audience, rather than to the broader public at large. With the advent of new media, this definition as been updated (see, for instance,~\cite{hirst2014communication}). In fact, from the perspective of readers of online content, they can decide whether and how to share such content by either broadcasting it to their audience or to address only a small portion of it (narrowcasting). In~\cite{virality}, for example, the act of forwarding a news article by email is treated as a form of narrowcasting, since in such case subjects are addressing a selected section of their audience/contacts.

Following the above definition, we will consider article comments as a form of \emph{narrowcasting}, as the readers who upload a comment on the article page are contributing to a discussion which 
happens at most between the readers of that article (and most probably, in fact, to a small subset of it). On the other hand, we consider the act of sharing an article to social 
networking sites as a form of \emph{broadcasting}. 

The rationale behind this distinction of viral facets in narrowcasting and broadcasting lies in previous research efforts showing that virality can be assessed through  
different metrics which represent distinct phenomena: it is not only the magnitude of spreading (virality) that depends on content but also the 
users reactions (comments, shares, tweets, etc.)~\cite{guerini2012linguistic,guerini2013exploring,shuai2012scientific}. 

In the following sections, all virality indices have been standardized in order to make them comparable ($\mu=0, \sigma=1$) across the two datasets. Emotion scores range between 0 and 1.


\section{Emotion Analysis}
\label{sec:emo}

To analyze the relationship between an article emotional characteristics and the corresponding impact on virality indices we use simple linear models. In this section we will discuss the importance of each emotion for the various virality indices, while the overall explanatory power of the models will be discussed in Section \ref{sec:r_square}.

Results on the \texttt{rappler.com} dataset, shown in Table~\ref{tab:rappler_res}, are in accordance with the findings of Berger et al.~\cite{virality}: high influence of \textsc{inspiring} 
(awe) and of negative-valence and high-arousal emotions such as \textsc{anger}, jointly with a low influence of \textsc{sadness}.
This similarity stands out for the specific case of \emph{narrowcasting}, and it is maintained for \emph{broadcasting}, albeit to a lower extent.

As mentioned in Section~\ref{sec:relwork}, other previous research works~\cite{fan2013anger,hochreiter2014role}, focusing on diverse scenarios and using heterogeneous datasets, have reported results consistent with these findings.

Nonetheless, turning to the Italian resource used in our analyses, we notice that results obtained from the \texttt{corriere.it} data, summarized in Table~\ref{tab:corriere_res}, are partially in line 
with~\cite{virality} for what concerns \emph{narrowcasting}, while drastically diverge on \emph{broadcasting}: in the latter case, \textsc{sadness} (a low-arousal 
emotional dimension) is found to be most relevant for virality, disproving the hypothesis that arousal alone can be used to explain virality phenomena. It should be noted that the lack of an \textsc{inspired} dimension for \texttt{corriere.it} cannot account for such discrepancy on \emph{broadcasting}, since it seems very unlikely that its potential votes would have been collected by \textsc{sadness}. 


It can be hypothesized, as a consequence, that strong cultural differences emerge when it comes to emotions (more precisely, to explicitly tagging one's own feeling on a website). 
What factors might underlie this phenomenon (e.g. historical period, editorial choices, deeper cultural sensibilities, etc.) represents a very fascinating research question, which 
is out of the scope of this work. 


Rather than focusing on specific emotions, in the next section we attempt to provide a more general explanation.

\begin{table} [!ht] 	
	\begin{center} 	 	
		{		
			\begin{tabular}{l|rrr} 		
				\hline 
 EMOTION & Valence & Arousal & Dominance\\
 \hline 
AFRAID & 2.25 & 5.12 & 2.71\\
AMUSED & 7.05 & 4.27 & 5.93\\
ANGRY & 2.53 & 6.02 & 4.11\\
ANNOYED & 2.80 & 5.29 & 4.08\\
DONT\_CARE & 3.53 & 4.27 & 3.62 \\
HAPPY & 8.47 & 6.05 & 7.21\\
INSPIRED & 6.89 & 5.56 & 7.30\\
SAD & 2.10 & 3.49 & 3.84\\
\hline 
			\end{tabular} 		
		} 		 	
	\end{center}	 	
	\caption{Valence, Arousal and Dominance scores for emotion labels, as provided by~\cite{warriner2013norms}.} 	
	\label{tab:VAD_scores} 
\end{table}


\begin{table*}[!ht]
\centering
{ 	
\begin{minipage}[b]{.45\textwidth}
\begin{tabular}{lrrr}
\multicolumn{4}{c}{\texttt{corriere.it} comments} \\
  \hline
 & Estimate & Std. Err& Sigf. \\ 
  \hline
  AROUSAL & .3741 & .0372 & *** \\ 
  VALENCE & -.3155 & .0486 & *** \\ 
  DOMINANCE & .1000 & .0761 &  $\dagger$ \\ 
   \hline
\end{tabular}
\end{minipage}
\begin{minipage}[b]{.45\textwidth}
\begin{tabular}{lrrr}
\multicolumn{4}{c}{\texttt{rappler.com} comments} \\
  \hline
 & Estimate & Std. Err& Sigf. \\ 
  \hline
  AROUSAL & .1205 & .0101 & *** \\ 
  VALENCE & -.1636 & .0165 & *** \\ 
  DOMINANCE & .0728 & .0221 & ** \\ 
   \hline
\end{tabular}
\end{minipage}

\begin{minipage}[b]{.45\textwidth}
\begin{tabular}{lrrr}
\multicolumn{4}{c}{\texttt{corriere.it} tweets} \\
  \hline
 & Estimate & Std. Err & Sigf. \\ 
  \hline
  DOMINANCE & .2158 & .0709 & ** \\ 
  VALENCE & -.2176 & .0453 & *** \\ 
  AROUSAL & .0400 & .0348 &  $\dagger$ \\ 
  
   \hline
\end{tabular}
\end{minipage}
\begin{minipage}[b]{.45\textwidth}
\begin{tabular}{ lrrr}
\multicolumn{4}{c}{\texttt{rappler.com} tweets} \\
  \hline
 & Estimate & Std. Err & Sigf. \\ 
  \hline
  DOMINANCE & .2244 & .0221 & *** \\ 
  VALENCE & -.1495 & .0165 & *** \\ 
  AROUSAL & .0065 & .0101 &  $\dagger$ \\ 
   \hline
\end{tabular}
\end{minipage}

\begin{minipage}[!h]{.45\textwidth}
\begin{tabular}{ lrrr}
\multicolumn{4}{c}{\texttt{corriere.it} g$+$ shares}\\
  \hline
 & Estimate & Std. Err & Sigf. \\ 
  \hline
  DOMINANCE & .4817 & .0704 & *** \\ 
  VALENCE & -.3781 & .0450 & *** \\ 
    AROUSAL & -.0242 & .0345 &  $\dagger$ \\ 
   \hline
\end{tabular}
\end{minipage} 
\begin{minipage}[b]{.45\textwidth}
\begin{tabular}{ lrrr}
\multicolumn{4}{c}{\texttt{rappler.com} g$+$ shares} \\
  \hline
 & Estimate & Std. Err & Sigf. \\ 
  \hline
  DOMINANCE & .2619 & .0221 & *** \\ 
  VALENCE & -.1530 & .0165  & *** \\ 
  AROUSAL & -.0371 & .0101 & *** \\ 
   \hline
\end{tabular}
\end{minipage}

}
\caption{Valence, Arousal and Dominance significant effects from simple Linear Models.} 
\label{tab:vad_results}
\end{table*}

\section{VAD analysis}
\label{sec:vad}
We now turn to investigate how basic constituents of emotions, such as Valence, Arousal, and Dominance (VAD), connect to virality.
The widely adopted VAD circumplex model of affect~\cite{bradley1994measuring,russell1980circumplex} maps emotions on a three-dimensional space, namely: Valence, denoting the 
degree of positive/negative affectivity (e.g. \textsc{fear} has high negative valence, while \textsc{joy} has high positive valence); Arousal, ranging from calming to exciting 
(e.g. \textsc{anger} is denoted by high arousal while \textsc{sadness} by low arousal); and Dominance, going from ``controlled'' to ``in control'' (e.g. \textsc{inspired}, highly 
in control, vs \textsc{fear}, overwhelming).

We thus proceed to map the emotions an article is found to evoke to the VAD circumplex model.
To do so, we exploit the work of Warriner et al.~\cite{warriner2013norms}, who provided a resource mapping roughly 14 thousands words to the VAD circumplex model, including
words representing the emotional dimensions we consider in this work -- see Table \ref{tab:VAD_scores}, which for us serve as a gold standard.
Such scores are in a Likert scale, ranging from 1 (low/negative) to 9 (high/positive).

Then, in order to compute VAD scores for a given article $doc$ we simply multiply the percentage of votes each emotion $e$ is found to evoke in $doc$ by the corresponding VAD score provided in Table \ref{tab:VAD_scores}, and then take the sum over the $n$ emotional dimensions considered. 
The formula for Valence is provided below; the equations used for Arousal and Dominance are akin.

\begin{equation}
doc_{v} = \sum\limits_{i=1}^n votes_\%(e_{i}) \times valence(e_{i})\end{equation}

As with virality indices, each VAD dimension has been standardized. 
Subsequently, we build linear models in order to assess whether and how VAD dimensions are connected to virality.
The results reported in Table~\ref{tab:vad_results} show very interesting trends:
VAD dimension estimates are found to be consistent among the two datasets (see estimate order and sign in Table~\ref{tab:vad_results}), in both \emph{broadcasting} 
(tweets/g+ shares) and 
\emph{narrowcasting} (comments) scenarios. Comparing these results with those provided in the previous sections, we see that the cross-cultural divergences in the relations between emotional dimensions and virality indices disappear when accounting for their more profound VAD components.

This finding hints at a generalized, culturally indipendent phenomenon: readers of the articles in our datasets tend to choose communication forms of \emph{narrowcasting} when the content is arousing but on which they 
feel less in control\footnote{Again, this result is in line with the intuition in~\cite{virality}, that arousal has a strong effect on \emph{narrowcasting}. Nonetheless, it appears of paramount importance to take the role of Dominance (not considered in~\cite{virality}) into account in order to understand \emph{broadcasting} phenomena.}.

Conversely, they turn to \emph{broadcasting} (e.g. share on social networks) when they feel more in control. This result is further supported by the fact that the less important dimensions (i.e. Dominance for  \emph{narrowcasting}, and Arousal for \emph{broadcasting}) are most of the time not significant, or with a slight negative impact on the model, indicating that these two dimensions \emph{switch roles} when transitioning from narrowcasting to broadcasting (and viceversa). 
This finding is important as it appears to be valid, in our analyses, among the two different languages (and cultural factors) our two datasets provide.

Finally, the role of Valence appears to be consistent among all indices of virality in both datasets, with negative valence contributing to higher virality. This result is in line with~\cite{hansen2011good}, who found that news-related content spreads more when imbued with negative Valence. 

%
%
%

\section{R$^2$ analysis}
\label{sec:r_square}
In this section we examine the explanatory power of our models and compare them with the results obtained by the models presented in Berger et al. \cite{virality} -- in particular, with \emph{model 2} (Positivity and Emotionality) and \emph{model 3} (Positivity and Emotionality plus the emotions Awe, Anger, Anxiety, Sadness). 

We compare our VAD models with \emph{model 2}, which was automatically computed starting from a lexicon of relevant affective words (the LIWC lexicon described in~\cite{Pennebaker99}), and our emotion-based models with \emph{model 3} in~\cite{virality}. It should be noted that the extensive work presented in~\cite{virality} includes other models accounting for variables (such as publication time, homepage position, author, among others) not available in our datasets. Nonetheless, it has been shown in~\cite{virality} that, while augmenting the explanatory power, these variables do not influence weight and role of the emotions in the models.



In~\cite{virality}, the dependent variable is represented as binary (i.e., 1 for those articles that make it on the \emph{most emailed list}, 0 for all the others). On the contrary, our viral 
indices are originally represented as continuous variables. In order to provide a fair comparison we thus proceeded to binarize the virality indices $V$ of article $n$, according to the following scheme:

 \begin{equation}
 V_n=\begin{cases}
    1, & \text{if $V_n > mean(V)+sd(V)$}\\
    0, & \text{otherwise}
  \end{cases}
 \end{equation} 

In this way, we obtain a small sample of the \emph{most viral} articles in our dataset, specular to the \emph{most emailed list} in~\cite{virality}. We also used logistic regression in accordance with~\cite{virality}. For comparison, we report in table~\ref{tab:R2_scores} the McFadden's R$^2$ used by Berger et al.~\cite{virality}, along with the standard R$^2$.

\begin{table} [!ht] 	
	\begin{center} 	 	
		{
			\begin{tabular}{l|rr} 		
				\hline 
 Viral Index & $R^2$ &  McFadden's $R^2$\\
\hline  
\multicolumn{3}{c}{\texttt{corriere.it} emotion models}  \\
\hline 
    Comments  & .0813 & .0922\\
    G+  & .0207 & .0527 \\
    Tweets  & .0084 & .0604 \\
\hline 
\multicolumn{3}{c}{\texttt{corriere.it} VAD models}  \\
\hline 
    Comments & .0530 & .0840  \\
    G+ & .0195 & .0446\\ 
    Tweets & .0072 & .0591\\
\hline 
\multicolumn{3}{c}{\texttt{rappler.com} emotion models}  \\
\hline 
    Comments  & .0191 & .0801\\
    G+  & .0115 & .0314 \\
    Tweets & .0112 & .0300 \\
\hline 
\multicolumn{3}{c}{\texttt{rappler.com} VAD models}  \\
\hline 
Comments & .0101 & .0569 \\ 
G+ & .0088 & .0288 \\
Tweets & .0100 & .0266 \\

 
\hline 
\multicolumn{3}{c}{Berger et al. models}  \\
\hline 
Model 2 & - & .0400 \\
Model 3 & - & .0700  \\
\hline 
			\end{tabular} 		
		} 		 	
	\end{center}	 	
	\caption{R$^2$ scores for the various linear models described and comparison with models presented in Berger et al.~\cite{virality}} 	
	\label{tab:R2_scores} 
\end{table}


Hence, when projecting emotions to their basic VAD dimensions, the models experience only a slight drop in terms of explanatory power, while gaining the advantage of language-invariance.

Moreover, our regression models for \emph{narrowcasting} have greater explanatory power than the best performing model in~\cite{virality}. This fact can be partially explained by the quality of our data: our datasets are much bigger and the affective annotations of the news articles were massively crowdsourced to the readers. 

Finally, 
emotions have significant effects on both \emph{narrowcasting} and \emph{broadcasting}, with a stronger impact on the former. Again, this finding is cross-cultural consistent.

\medskip

\section{Conclusions}
\label{sec:concl}

In this article we have provided a comprehensive investigation on the relations between virality of news articles and the emotions they are found to evoke. By exploiting a high-coverage and bilingual corpus of 65k documents containing metrics of their spread on social networks as well as a massive affective annotation provided by readers (more than 1.5 millions votes), we presented a thorough analysis of the interplay between evoked emotions and viral facets. 

We highlighted and discussed our findings in light of a cross-lingual approach.

Our results show differences in evoked emotions and corresponding viral effects across the bilingual data we crawled; amongst other findings, we note the remarkably discordant influence of the \textsc{sadness} affective dimension between the english- and italian- language datasets, in contrast with the hypothesis that arousal alone can be used to explain virality phenomena. 
While these findings seem to indicate effects of cultural differences on the relation existing between emotions and virality, such hypothesis could only be properly tested were more information on the annotators demographics available (e.g. geographic provenance).

Still, when accounting for the deeper constituents of emotions our analyses provided compelling evidence of a generalized 
explanatory model 
rooted in the Valence-Arousal-Dominance (VAD) circumplex. Viral facets seem to be coherently affected by particular VAD configurations (namely, the alternation between Dominance and Arousal), and these configurations indicate a clear connection with distinct phenomena underlying persuasive communication. In particular, high arousal is more connected to \emph{narrowcasting} phenomena, while Dominance to \emph{broadcasting} phenomena. 

Extensions of this work will include deeper investigations of the relations between VAD dimensions and virality. For instance, while in the analyses reported herein we have used a resource~\cite{warriner2013norms} to map affective tags to their VAD constituents, we plan to increase the resolution by associating each article with a VAD representation derived from its textual content.

The results presented in this article can be very valuable in contexts such as content marketing and native advertising, and thus be relevant not only for social science researchers interested in understanding the factors behind virality phenomena, but also for marketing and industry people.

\section*{Acknowledgements}
The work of M. Guerini has been supported by the Trento RISE PerTe project.
The work of J. Staiano has been partially funded by the French government through the National Agency for Research (ANR) under the Sorbonne Universit\'{e}s Excellence Initiative program (Idex, reference ANR-11-IDEX-0004-02) in the framework of the \emph{Investments for the future} program, and by the German Federal Ministry of Research and Education, within the EBITA project.

\bibliographystyle{abbrv}
\bibliography{Persuasive}
\end{document}